\documentclass[aps,twocolumn,prd,superscriptaddress,showpacs,preprintnumbers,amsmath,amssymb,nofootinbib]{revtex4-1}
\pdfoutput=1

\usepackage{hyphenat}
\usepackage[caption=false]{subfig}
\usepackage{graphicx}
\usepackage{enumitem}
\usepackage{mathtools}				
\usepackage{relsize}
\usepackage{xcolor}
\usepackage{dsfont}
\usepackage{units}					
\usepackage{csquotes}					
\usepackage{siunitx}
\usepackage{xspace}
\usepackage{multirow}
\usepackage{slashed}

\usepackage{hyperref}

\newcommand{\eg}{\emph{e.g.~}}
\newcommand{\cf}{cf.~}
\newcommand{\ie}{\emph{i.e.}~}
\newcommand{\CL}{CL\xspace}
\newcommand{\hc}{\text{h.c.\xspace}}
\newcommand{\br}{\mathcal{B}}
\newcommand{\ETmiss}{\slashed{E}_T}
\newcommand{\invHiggs}{\Delta_{mh}}
\newcommand{\invZ}{\Delta_{mZ}}
\newcommand{\recoilZ}{\Delta_\text{recoil}}
\newcommand{\DeltaPhi}[2]{\Delta\phi_{#1 #2}}
\newcommand{\DeltaR}[2]{\Delta R_{#1 #2}}

\newcommand{\I}{\mathrm{i}}				
\renewcommand{\Re}{\operatorname{Re}}	
\renewcommand{\Im}{\operatorname{Im}}	
\newcommand{\fineq}[1]{\;{#1}}				

\DeclarePairedDelimiter\absVal{\vert}{\vert}

\newcommand{\symhspace}[2]{\hspace{#1}#2\hspace{#1}}

\definecolor{darkblue}{rgb}{0.0,0.0,0.4}
\definecolor{darkgreen}{rgb}{0.0,0.4,0.0}
\hypersetup{
    colorlinks,
    linkcolor=darkblue,	
    citecolor=darkgreen,
    urlcolor=darkblue
}

\begin{document}


\title{Quark flavour-violating Higgs decays at the ILC}

\author{Daniele Barducci}
\email{daniele.barducci@sissa.it}
\affiliation{SISSA and INFN, Sezione di Trieste, via Bonomea 265, 34136 Trieste, Italy}

\author{Alexander J. Helmboldt}
\email{alexander.helmboldt@mpi-hd.mpg.de}
\affiliation{\vspace{0.5em}Max-Planck-Institut f\"ur Kernphysik, Saupfercheckweg 1, 69117
  Heidelberg, Germany}



\begin{abstract}
\noindent
Flavour-violating Higgs interactions are suppressed in the Standard Model  such that their observation would be a clear sign of new physics.
We investigate the prospects for detecting quark flavour-violating Higgs decays in the clean ILC environment.
Concentrating on the decay to a bottom and a light quark $j$, we identify the dominant Standard Model background channels as coming from hadronic Standard Model Higgs decays with mis-identified jets.
Therefore, good flavour tagging capabilities are essential to keep the background rate under control.
Through a simple cut-based analysis, we find that the most promising search channel is the two-jet plus missing energy signature $e^+e^-\to bj+\slashed{E}_T$.
At 500\,GeV, the expected 95\,\% CL upper limit on $\mathcal{B}(h\to bj)$ is of order $10^{-3}$.
Correspondingly, a $5\,\sigma$ discovery is expected to be possible for branching ratios as low as a few $10^{-3}$.
\end{abstract}

\begin{flushright}
\hspace{3cm} 
SISSA-50-2017-FISI
\end{flushright}

\maketitle


\section{Introduction}
\label{sec:intro}
\noindent
With the discovery of the Higgs boson at the Large Hadron Collider (LHC) in July 2012~\cite{Aad2012b,Chatrchyan2012a} the high energy particle physics program has entered a new exciting phase aiming at understanding whether the properties of the newly discovered state are consistent with the predictions of the Standard Model (SM). The 7 and \SI{8}{TeV} runs of the LHC have constrained the Higgs bosons couplings to $WW$, $ZZ$, $\gamma\gamma$ and $\tau\tau$ pairs to be approximately within \SI{10}{\%} of their SM values, while weaker limits have been derived for the interactions with bottom and top quarks as well as gluons~\cite{ATLAS2015}.

Obviously, more exotic Higgs couplings are of interest as well, in particular those non-diagonal in flavour space. Such interactions enable flavour-changing neutral currents (FCNC) and are thus forbidden at tree-level and, in the quark sector, GIM-suppressed at loop-level in the absence of new physics (NP). The corresponding processes in the SM are therefore very rare%
\footnote{In the SM one has for example ${\cal B}(h\to b \bar s)\sim\num{e-7}$~\cite{Bejar:2004rz}.}
and beyond the reach of  today's experiments. Hence any observation would constitute a clear sign of physics beyond the SM. Indeed, there are various well-motivated models which predict sizable flavour-violating Higgs couplings, such as two Higgs doublet models~\cite{DiazCruz:1999xe,Branco:2011iw,Crivellin:2013wna,Botella:2015hoa,Crivellin:2017upt}, supersymmetric models~\cite{Arhrib:2012ax}, models with a composite Higgs~\cite{Agashe:2009di}, models of extra dimensions~\cite{Azatov:2009na} and others.

Consequently, the ATLAS and CMS collaborations have put forward a program aiming at the search for such couplings. In particular, flavour-violating Higgs decays in the leptonic sector have received great attention due to possible indications of a non-zero \mbox{$h\to \tau \mu$} branching ratio in \SI{8}{TeV} data~\cite{Khachatryan2015}, which was, however, not confirmed by more recent analyses performed with the first tenths of inverse femtobarn of integrated luminosity accumulated with \SI{13}{TeV} proton-proton collisions~\cite{CMS:2017onh}.
In contrast, the situation is somewhat more complicated in the quark sector.
On the one hand, flavour-violating Higgs couplings involving a top quark can be tested either via exotic top quark decays~\cite{Aad:2015pja,Khachatryan:2016atv} or in principle, given the mass of the Higgs boson, via Higgs decays to an off-shell top, a process which however seems to be outside the LHC reach.
On the other hand, processes involving bottom quarks suffer from a huge QCD background which makes a measurement of decays such as \mbox{$h\to b j$}, with $j$ representing a light quark, extremely challenging at the LHC, even at the end of the high luminosity phase when ${\cal O}(\SI{3000}{fb^{-1}})$ of data will have been collected.

It is thus crucial to investigate whether proposed future colliders could offer a handle to \textit{directly} test processes for which the LHC has a poor sensitivity.
Direct probes of such decays are in fact essential both to complement indirect limits on FCNC interactions arising from low-energy flavour observable measurements and in case of deviations from the SM expectations being observed to be able to pin down their origin.
Correspondingly, it is the aim of this paper to study the potential of an $e^+e^-$ collider in probing exotic decay modes of the Higgs boson in a bottom quark and a light quark. In doing so, we will focus on the planned International Linear Collider (ILC), adopting the proposed centre-of-mass energy, beam polarisation and integrated luminosity values of this machine~\cite{Behnke2013,Baer2013} and showing that the sensitivity on NP couplings responsible for \mbox{$h\to b j$} processes competes with, and for some specific analysis surpasses, the one that can be obtained by low-energy flavour measurements in the effective field theory (EFT) approach.

The paper is organised as follows. In Sec.~\ref{sec:param} we describe the effective parametrisation that we use for our study while in Sec.~\ref{sec:HiggsILC} we review the Higgs physics program at the ILC. In Sec.~\ref{sec:results} we describe the details of our numerical analysis and present our findings. We then conclude in Sec.~\ref{sec:conc}.

\section{Parametrisation}
\label{sec:param}
\noindent
We chose to parametrise the non-SM interactions of the Higgs boson with a pair of quarks using the EFT language adopting as a basis the one of the Warsaw convention~\cite{Grzadkowski:2010es}. At dimension-six level, the operators that couple the Higgs field with a pair of down-type quarks are%
\begin{subequations}
	\label{eq:d6}
	 \begin{align}
		{\cal Q}_{d H} & = (\bar q^i_L H d^j_R) (H^\dag H)\label{eq:d6:1},\\
		{\cal Q}_{H d} & = (\bar d^i_R \gamma^\mu d^j_R)(H^\dag \I \overleftrightarrow{D_\mu} H), \label{eq:d6:2}
	 \end{align}
\end{subequations}
where $q_L$ and $d_R$ indicate the left-handed doublet and right-handed singlet of SM quarks with generation indices $i,j$, $H=(\phi^+,v+h)^\intercal/\sqrt{2}$ is the Higgs field with $v\simeq246$ GeV, $h$ the physical Higgs state and $H^\dag \I \overleftrightarrow{D_\mu} H=H^\dag \I D_\mu H - (\I D_\mu H^\dag)H$. However, after electroweak symmetry breaking (EWSB) only the operator of Eq.~\eqref{eq:d6:1} gives rise to interactions of the physical Higgs with a pair of SM quarks of the form
 \begin{equation}
 {\cal L_{\rm EFT}} = -y_{ij} h d^i_L d^j_R + \hc \label{eq:d6:3}
 \end{equation}
These interactions have numerous potentially observable effects. Firstly, they introduce additional decay modes for the Higgs boson, with a leading order partial width in the limit $m_{d^i},m_{d^j} \ll m_h$ given by
\begin{align}
	\Gamma(h\to d^i d^j) = \frac{N_c m_h}{8\pi} \bigl( \absVal{y_{ij}}^2 + \absVal{y_{ji}}^2 \bigr) \fineq{.}
	\label{eq:decayrate}
\end{align}

Secondly, they generate FCNC at tree-level, which are testable in $K^0$ and $B^0_{s,d}$ meson oscillation experiments.
Flavour measurements impose stringent limits on the values of the Wilson coefficient of Eq.~\eqref{eq:d6:3}, which are reported in Tab.~\ref{tab:flavour-limits}.
These bounds in turn translate into upper limits on the exotic Higgs branching ratios. For example, assuming non-zero real Wilson coefficients only for the $bs$ sector one gets $\br(h\to bs)<\num{1.8e-3}$ for $y_{bs}=y_{sb}$ and $\br(h\to bs)<\num{6.8e-3}$ with $y_{bs}\ne 0$ and $y_{sb}=0$ or vice versa.
However, given the unspecified nature of the ultraviolet structure of the model captured by the EFT framework, there could exist in principle 
additional effects giving rise to cancellations in the Feynman diagrams leading to mesons oscillations, still allowing for a sizable $h\to b s$ decay rate, see {\emph{e.g.}}~the recent~\cite{Crivellin:2017upt}.
For this reason we will consider the exotic Higgs branching ratio as a free parameters in the following.

\begin{table}[h]
	\centering
	\begin{tabular}{c | c | c}
    \toprule
    \symhspace{1.2em}{Observable} & \symhspace{1.7em}{Coupling} & \symhspace{3.0em}{Constraint}              \\ 
   \colrule
    \multirow{2}{*}{$B_d^0$ oscillations}
                  & $|y_{db}|^2$, $|y_{bd}|^2$             &  $<\num{2.3e-8}$ \\
                  & $|y_{db} y_{bd}|$                      &  $<\num{3.3e-9}$ \\ \colrule
     \multirow{2}{*}{$B_s^0$ oscillations}
                  & $|y_{sb}|^2$, $|y_{bs}|^2$             &  $<\num{1.8e-6}$ \\
                  & $|y_{sb} y_{bs}|$                      &  $< \num{2.5e-7}$ \\ \colrule
     \multirow{4}{*}{$K^0$ oscillations}
                  & $\Re y_{ds}^2$, $\Re y_{sd}^2$       &  $[-5.9 \dots 5.6] \cdot 10^{-10}$ \\
                  & $\Im y_{ds}^2$, $\Im y_{sd}^2$       &  $[-2.9 \dots 1.6] \cdot 10^{-12}$ \\
                  & $\Re y_{ds}^* y_{sd}$                 &  $[-5.6 \dots 5.6] \cdot 10^{-11}$ \\
                  & $\Im y_{ds}^* y_{sd}$                 &  $[-1.4 \dots 2.8] \cdot 10^{-13}$ \\ 
                  \botrule
	\end{tabular}
  \caption{Current \SI{95}{\%} confidence level (\CL) limits on the Wilson coefficient of Eq.~\eqref{eq:d6:3} from meson oscillations experiments.
    The reported bounds are taken from~\cite{Harnik2013}.}
  \label{tab:flavour-limits}
\end{table}


\section{Higgs physics at the ILC}
\label{sec:HiggsILC}
\noindent
One of the ILC's major physics goals is the precise measurement of the Higgs bosons' properties (for a review, see \eg \cite{Asner2013}). Thus, not only a deeper insight into the mechanism of EWSB will be acquired, but also NP that couples to the Higgs may be detected directly or indirectly.
In particular, search channels suffering from large QCD backgrounds which are hard or impossible to see at the LHC may become easily accessible in the clean collider environment of the ILC. 
This is for example the case of the searches for the Higgs boson decaying into a pair of bottom quarks, for which only recently the ATLAS and CMS collaborations have reported a direct evidence of approximately $\SI{3.5}{\sigma}$~\cite{CMS:2017tkf,Aaboud:2017xsd}. Needless to say, Higgs decays into a bottom and a light quark are even more challenging and possibly beyond the LHC reach.

In order to have a sufficient number of Higgs bosons to be investigated, all ILC operating scenarios include runs at a centre-of-mass energy of \SI{250}{GeV} where the Higgs-strahlung production cross section, \mbox{$e^+e^-\to Zh$}, reaches its maximum of around \SI{300}{fb}.%
\footnote{All cross-sections quoted in this section are valid assuming polarised beams with $\mathcal{P}(e^-,e^+) = (-0.8,+0.3)$, see \eg \cite{Baer2013}.}
Hence, different event signatures are possible depending on the actual $Z$ decay taking place. Of course, all of them come with their own advantages and drawbacks.

On the one hand, due to the large branching ratio $\br(Z\to \bar{q}q)\simeq\SI{70}{\%}$,%
\footnote{For our analysis we adopt the $Z$ branching ratios reported in~\cite{Patrignani2016}.} events with a hadronically decaying $Z$ are the most abundant ones.
Unfortunately, they are also the least clean ones. In particular, as in our case the signal process involves the Higgs decaying into quarks as well, the arising final state contains four jets.
This signature comes along with a large SM background both due to processes where a Higgs boson decays into a pair of same flavour jets of which at least one flavour is mis-identified, and to four-jet production which, despite presenting a different final state kinematics, have a large production cross section. 
Still, we will investigate the potential of the hadronic search channel in Section \ref{sec:results:hadronic}.

On the other hand, the $Z$ may also decay into electron or muon pairs. These particles can be reconstructed and identified with good efficiencies. Thereby, it is possible to very reliably tag the produced Higgs via the recoil mass technique which allows to precisely measure the mass of the Higgs boson~\cite{Baer2013} and to reject backgrounds arising from processes without its resonant propagation.
However, the corresponding $Z$ branching ratio is roughly an order of magnitude smaller than in the hadronic case, $\br(Z\to\ell^+\ell^-) = \br(Z\to e^+e^-) + \br(Z\to \mu^+\mu^-)\simeq\SI{6.7}{\%}$, thus possibly limiting the statistical significance that can be reached with this search mode.
The analysis of this channel is presented in Section \ref{sec:results:ll250}.

Finally, the $Z$ can decay invisibly into neutrinos with a branching ratio of $\br(Z\to \bar{\nu}\nu)\simeq\SI{20}{\%}$.
The signal signature then contains missing transverse energy%
\footnote{Given the precise knowledge about the colliding particles' energies, we could employ the total missing momentum instead of $\ETmiss$, the latter being widely used in hadron collider studies. We expect, however, that due to the uncertainties resulting from a non-trivial beam energy spectrum, beamstrahlung and initial state radiation, this will not improve upon the results obtained.}
($\ETmiss$) that can be used to select the event along with the information from the Higgs decay products. However, the presence of missing energy implies that more background channels have to be taken into account since the $bj + \ETmiss$ signature can also arise, {\emph{e.g.}}, through $\ell \nu c s$ final states where the lepton remains undetected and the charm quark is mis-identified.
In any case the neutrino channel turns out to be a good compromise between event rate and cleanliness and will be covered in Section \ref{sec:results:vv}.

All other Higgs production processes only give minor contributions at $\sqrt{s}=\SI{250}{GeV}$.
In contrast, at a centre-of-mass energy of \SI{500}{GeV} most Higgs bosons are produced in $W$-fusion with a cross-section of approximately \SI{160}{fb}.
Obviously, this results in a signature very similar to the one from Higgs-strahlung with subsequent invisible $Z$ decays, the cross-section of which is around \SI{20}{fb} at this energy.
We will investigate the reach of the neutrino search channel during the \SI{500}{GeV} run in Section \ref{sec:results:vv}.


\section{Analysis and results}
\label{sec:results}

\subsection{Analysis details}
\noindent
In order to perform a detector-level analysis, we first employed the {\tt FeynRules} package~\cite{Alloul:2013bka} to implement the effective vertices of Eq.~\eqref{eq:d6:3} in the {\tt UFO} format~\cite{Degrande:2011ua}.
Events were then generated using {\tt Whizard 2.5.0}~\cite{Kilian2011} supplemented by the {\tt O'Mega} matrix element generator~\cite{Moretti2001}. In doing so, polarised beams and a non-trivial beam energy spectrum due to initial state radiation and beamstrahlung were taken into account.
Results without the inclusion of the aforementioned effects were eventually cross checked with {\tt MadGraph5\_aMC@NLO}~\cite{Alwall:2014hca}.
Parton showering and hadronisation were performed through {\tt PYTHIA8}~\cite{Sjoestrand2015}, while a fast detector simulation was carried out using {\tt Delphes 3}~\cite{Favereau2014} supplemented by the {\tt DSiD} detector card suitable for performing analyses at $e^+e^-$ linear colliders~\cite{Potter2016b}.
Jets were reconstructed with {\tt FastJet}~\cite{Cacciari2012} via the {\tt anti-kT} algorithm~\cite{Cacciari2008b} with a cone radius of \num{0.4}.
Events have been finally analysed with {\tt MadAnalysis5}~\cite{Conte:2012fm}.

For the purpose of accurately modeling the flavour identification of $b$ and $c$ jets at the ILC, we assumed the following tagging efficiencies
\begin{subequations}
\label{eq:tag}
\begin{align}
	\mathcal{P}(b\text{-tag}\,|\,b) & = 0.80 \,,\quad \mathcal{P}(c\text{-tag}\,|\,c) = 0.70\,,
	\label{eq:tag:eff}
\intertext{and mis-identification rates}
	\begin{split}
		\mathcal{P}(b\text{-tag}\,|\,c) & = 0.08 \,,\quad \mathcal{P}(b\text{-tag}\,|\,j) = 0.01\,, \\
		\mathcal{P}(c\text{-tag}\,|\,b) & = 0.17 \,,\quad \mathcal{P}(c\text{-tag}\,|\,j) = 0.10\,,
	\end{split}
	\label{eq:tag:misID}
\end{align}
\end{subequations}
which are based on the expected performance of the {\tt LCFIPlus} software~\cite{Suehara2016}.
Note that at the ILC it will not be possible to differentiate between jets from $s$ and $d$ quarks, rendering the Higgs decays to $b s$ and $b d$ effectively indistinguishable.
Hence, our analysis is sensitive to the quantity $\br(h\to bj):=\br(h\to b s)+\br(h\to b d)$ rather than to the individual branching ratios.

One of the major advantages of a linear collider is the ability to polarise its particle beams. For the ILC, beam polarisations of \mbox{${\cal P}(e^-)=\pm \SI{80}{\%}$} and \mbox{${\cal P}(e^+)=\pm \SI{30}{\%}$} are expected. During our analyses we will, on the one hand, compare the performance of the four possible combinations and, on the other hand, we will consider two potential ways of how to realistically split the full integrated luminosity among the combinations in an actual run, namely~\cite{Barklow2015}
\begin{align*}
	\text{\bf Scenario 1:}& \quad \mathcal{P}_\mathsmaller{-+,+-,++,--}=[\SI{67.5}{\%}, \SI{22.5}{\%}, \SI{5}{\%}, \SI{5}{\%}] \,, \\
	\text{\bf Scenario 2:}& \quad \mathcal{P}_\mathsmaller{-+,+-,++,--}=[\SI{40}{\%}, \SI{40}{\%}, \SI{10}{\%}, \SI{10}{\%}] \,,
\end{align*}
where the first (second) sign corresponds to the polarisation of the electron (positron) beam.
Regarding the amount of accumulated data, we adopt the specifications of the H-20 scenario, \ie we assume an integrated luminosity of $\SI{2}{ab^{-1}}$ ($\SI{4}{ab^{-1}}$) at a centre-of-mass energy of \SI{250}{GeV} (\SI{500}{GeV}), with an initial phase at both energies collecting $\SI{500}{fb^{-1}}$ each~\cite{Barklow2015}.

For later reference, we define the signal significance as
\begin{equation}
	z=\frac{S}{\sqrt{S+B+ \epsilon_{\text{syst}}^2 B^2}} \,,
	\label{eq:gaussZ}
\end{equation}
where $S$ and $B$ indicate the number of signal and background events, respectively, and $\epsilon_\text{syst}$ is the systematic uncertainty on the background determination, \ie $\epsilon_{\text{syst}}=\Delta B/B$.
For given $B$ and $\epsilon_\text{syst}$, the \SI{95}{\%} \CL upper limit on the number of signal events is then defined as the value $S_{95\%}$, for which Eq.~\eqref{eq:gaussZ} gives $z=\Phi^{-1}(0.95)\approx1.64$, where $\Phi^{-1}$ is the probit function.

The remaining sections will be divided in line with our discussion in Section \ref{sec:HiggsILC}, according to signal signatures and centre-of-mass energies.
Whenever an explicit rate for the signal process is of relevance, we choose as benchmark point $\absVal{y_{bs}}=\absVal{y_{sb}}=\num{e-3}$, which gives an exotic branching fraction of $\br(h\to bs)\simeq\SI{0.73}{\%}$ according to Eq.~\eqref{eq:decayrate}.


\subsection{Hadronic channel at 250\,GeV}
\label{sec:results:hadronic}
\noindent
The SM background for the hadronic channel arises from the inclusive production of four jets, \mbox{$e^+ e^- \to k k k k$}, where $k$ indicates gluons or any quark excluding the top.
We have split this background into the following two contributions:
\begin{description}[before={\renewcommand\makelabel[1]{\upshape\bfseries ##1}}]
	\item[Non-resonant processes] where the Higgs boson does not propagate in the Feynman diagrams (hereafter referred to as $4k$).
	\item[Resonant Higgs processes] where the Higgs is produced on-shell in association with two quarks and then further decays into a pair of jets ($\bar{b}b, \bar{c}c, gg)$. This class includes both Higgs-strahlung type processes with a hadronically decaying $Z$ and topologies with quark pair production, where one quark radiates off a Higgs boson.
\end{description}
The above splitting has been performed in order to be able to rescale the final number of events in order to properly take into account the SM Higgs branching ratios as predicted by the Higgs cross section working group for a \SI{125}{GeV} Higgs boson~\cite{deFlorian:2016spz}.
Note that this procedure removes the contribution to the background given by the interference amongst the two classes of processes which, however, we found to be negligible.

\begin{figure}[!h!]
    \centering
            \includegraphics[scale=0.83]{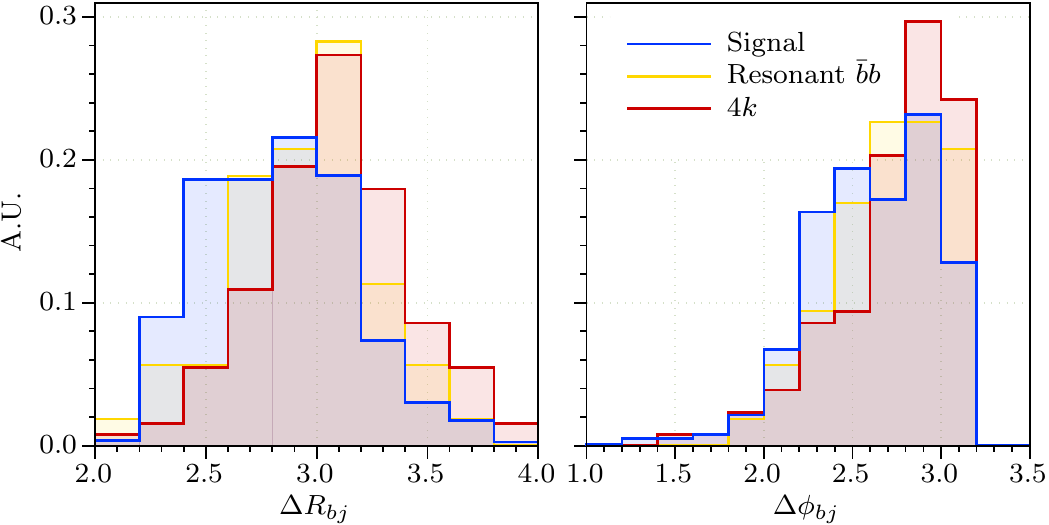}\hfill
    \caption{Normalised distributions of the $\Delta R$ and $\Delta\phi$ separation between the  $b$ jet and the light jet reconstructing the Higgs mass after the invariant mass cuts for $\sqrt{s}=\SI{250}{GeV}$ with polarisation ${\cal{P}}_{-+}$ for both the signal and the two main background processes.}
    \label{hadr:distr-hadr}
\end{figure}

Signal events are selected with the following requirements. We ask for exactly four jets, with at least one $b$-tag jet and one light-flavour jet.
All jets are supposed to satisfy $p_T>\SI{10}{GeV}$ and $\absVal{\eta}<1.8$.
We further require that a $b$-tag jet and a light jet reconstruct the Higgs mass within $\invHiggs:=\absVal{m_{bj}-m_h}<\SI{5}{GeV}$ and that two same-flavour jets, different from the ones identifying the Higgs mass, reconstruct the $Z$ boson mass with $\invZ:=\absVal{m_{kk}-m_Z}<\SI{10}{GeV}$.
In case that more than one pair falls into the chosen invariant mass window, we chose the pair closest to the true mass.
We finally optimise the signal selection by asking the $b$-jet arising from the Higgs boson decay to have a transverse momentum greater than \SI{50}{GeV} and a not too large angular separation from the light jet with which it reconstructs the Higgs mass: $\DeltaR{b}{j}<2.6$ and $\DeltaPhi{b}{j}<2.5$.
These last two selection cuts have been imposed to further suppress the combinatorial $4k$ background.
In Fig.~\ref{hadr:distr-hadr}, we show the differential distributions for the two variables normalised to a unitary area after the invariant mass cuts which motivates our angular requirements.

\begin{figure}[h!]
	\centering
	\includegraphics[scale=0.838]{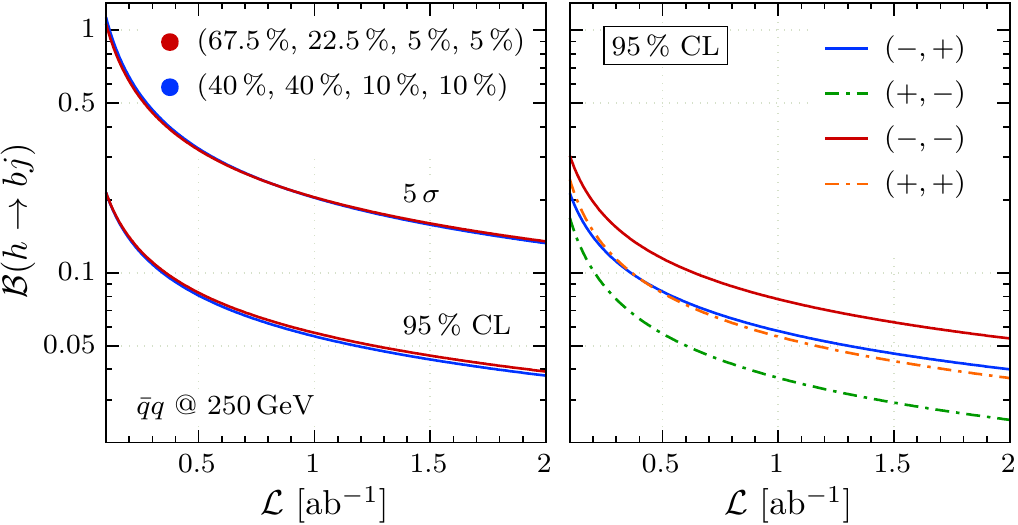}
	\caption{{\emph{Left}}: Expected \SI{95}{\%} \CL exclusion and $\SI{5}{\sigma}$ discovery reaches on \mbox{${\cal{B}}(h\to bj)$} in the hadronic channel at $\sqrt{s}=\SI{250}{GeV}$ as a function of the total integrated luminosity for the two polarisation scenarios described in the text. {\emph{Right}}: Expected exclusion reaches for the four possible polarisation configurations individually.  Both plots assume $\epsilon_\text{syst}=\SI{1}{\%}$.}
	\label{fig:hadr}
\end{figure}

For the above choices of selection cuts the event yields for the signal and background processes for polarisation scenario 1 and with an integrated luminosity of $\SI{2}{ab^{-1}}$ are reported in Tab.~\ref{tab:had:cutflow} for our benchmark point with ${\cal{B}}(h\to bs)\simeq \SI{0.73}{\%}$.
Assuming a systematic background error of \SI{1}{\%}, we then show in Fig.~\ref{fig:hadr} the \SI{95}{\%}~CL and $\SI{5}{\sigma}$ contours in the luminosity-branching ratio plane for all the polarisation configurations individually as well as for the two sharing scenarios mentioned before.
The hadronic channel will thus be able to exclude at \mbox{\SI{95}{\%} \CL} branching ratios ${\cal{B}}(h\to b j)$ of around $\SI{5}{\%}$ with $\SI{2}{ab^{-1}}$ of accumulated data for both realistic polarisation sharing scenarios.
A bound lower by a factor of two could in principle be obtained with the ILC running in the ${\cal{P}}_{+-}$ configuration up to the full targeted integrated luminosity at $\sqrt{s}=\SI{250}{GeV}$.

\begin{table}[h]
	\centering
	\begin{tabular}{l | cccccc}
		\toprule
		& {Signal} & {$4k$} & {$\bar b b$} & {$\bar c c$} & {$gg$} \\
		\colrule		
		Exp. 									& 2681 	& \num{2.8e7} & \num{1.9e5} & 9490 & \num{2.8e4}\\		
		Jet tag. 							& 1080 	& \num{1.9e6} & \num{6.9e4} &1150 & 3470\\		
		$\invHiggs$ 						& 147 	& \num{1.4e5} & 6730 					&87 & 223\\		
		$\invZ$    							& 21 		& 6946					   & 98 						&5 & 2\\		
		$p_T^b$							& 14 		& 3503					   & 81 						& 3& 1\\		
		$\Delta{\phi_{bj}}$ 			& 5 		& 419					   & 15 						& 1& 1\\		
		$\Delta{R_{bj}}$					& 4 		& 165					   & 12						 &1 & 0\\	
		\botrule
	\end{tabular}
	\caption{Cutflow table for the signal and backgrounds in the hadronic channel at \SI{250}{GeV} with polarisation sharing scenario 1 and $\SI{2}{ab^{-1}}$ of integrated luminosity. For the signal we have assumed $\br(h\to bj)\simeq \SI{0.73}{\%}$.}
\label{tab:had:cutflow}
\end{table}

\subsection{Leptonic channels}
\label{sec:results:leptonic}
\noindent
The previous section demonstrated that the reach of the hadronic channel is limited due to the large combinatorial four-jet background.
If we want the signal signature to contain only a pair of jets there are two possibilities (\cf Section \ref{sec:HiggsILC}).
Either the final state involves a pair of opposite-sign charged leptons from $Z$ decays (\textit{charged lepton channel}), or there is missing energy coming from a pair of neutrinos (\textit{neutrino channel}).
Clearly, both signatures come with different kinds of backgrounds, that we discuss in detail in the following subsections.

At this point, we only mention that the dominant SM background for both leptonic channels comes from on-shell Higgs production with its subsequent decay into a pair of same-flavour quarks. Due to flavour tagging imperfections, these processes can mimic the signal, while, in particular, exhibiting its identical kinematics. They are thus \textit{de facto} irreducible and are only suppressed by flavour-tagging requirements.
It is therefore simple, yet very instructive,%
\footnote{For the hadronic channel such an estimate is less useful due to the large amount of combinatorial background.}
to estimate the relative importance of the individual background processes based only on the SM Higgs branching ratios and the flavour tagging specifications from Eq.~\eqref{eq:tag}.
The results are compiled in Table \ref{tab:bkgdEstimate}, where the relative weight measure $r$ is defined as
\begin{align*}
	r_X = \frac{\left. \br\cdot \epsilon_\text{tag} \right|_X}{\left. \br\cdot \epsilon_\text{tag} \right|_{bj}} \,
\end{align*}
with $\epsilon_\text{tag}$ being the probability that the given quark pair is tagged as $bj$. Consequently, the resonant $\bar{b}b$ process is expected to be by far the most severe background.

\begin{table}[h!]
	\centering
	\begin{tabular}{l|ccc}
		\toprule
		Process $X$\hspace{0.7em} & {\symhspace{0.5em}{$\br(h\to X)$ [\%]}} & {\symhspace{0.5em}{$\epsilon_\text{tag}$ [\%]}} & {\hspace{0.5em}relative weight $r$} \\
		\colrule
		\quad$bj$ & 0.73 & 71.2 & 1.0 \\
		\quad$bb$ & 58.09 & 4.8 & 5.4 \\
		\quad$cc$ & 2.884 & 3.52 & 0.20 \\
		\quad$gg$ & 8.180 & 1.78 & 0.28 \\
		\botrule
	\end{tabular}
	\caption{Qualitative estimate of the irreducible backgrounds' relative importance for the leptonic search channels. Higgs branching ratios $\br$ are taken from \cite{deFlorian:2016spz}, while tagging efficiencies $\epsilon_\text{tag}$ are calculated based on Eq.~\eqref{eq:tag}.}
	\label{tab:bkgdEstimate}
\end{table}

Based on the above considerations, we can even obtain a qualitative estimate of the reachable statistical significance $z$ given the expected number of signal events $S$ and the sum of all background relative weights $r_B$, namely $z \approx \sqrt{S/(1+r_B)}$.
For instance, plugging in the numbers from Tab.~\ref{tab:bkgdEstimate} yields $ \sqrt{S}/\num{2.6}$.
Clearly this is just a very rough upper bound for $z$ due to several reasons including the fact that we took into account only the resonant background modes and ignored systematic errors. However, it helps us stress the importance of good flavour tagging for a successful analysis once more.
Let us, {\emph{e.g.}},~assume the same specifications for $b$ tagging as before, but a degraded $c$ tagging efficiency of only \SI{60}{\%}. Even though the mis-tag rates will then decrease accordingly (subject to the \texttt{LCFIPlus} performance \cite{Suehara2016}), redoing the above calculation shows a further enhancement of the $\bar{b}{b}$ background, $r_{bb}=\num{13.5}$, with an expected worsened  significance $z\approx\sqrt{S}/\num{3.9}$.

Note that the above estimates were obtained for a given fixed branching ratio \mbox{$\br(h\to bj)=\SI{0.73}{\%}$}.
Alternatively, we can ask for the minimally achievable \SI{95}{\%} CL upper bound on $\br$. Again taking into account only resonant backgrounds and neglecting systematic errors, we obtain from Eq.~\eqref{eq:gaussZ}
\begin{align}
	\br_{95\%} \gtrsim \num{1.3}\cdot\br_0/S_0 \left( 1 + \sqrt{1 + \num{0.063}\cdot S_0/\br_0} \right) \,,
	\label{eq:bkgdEstimate:br95}
\end{align}
where we used the SM Higgs branching ratios and tagging efficiencies of Table \ref{tab:bkgdEstimate}.
Notice that $\br_0$ is some reference signal branching fraction while $S_0$ is the corresponding number of observable events.
Hence, the ratio $S_0/\br_0$ encodes the performance of a given analysis setup and is not sensitive to any BSM parameter.
In Section \ref{sec:results:vv}, we will compare the results of our actual analysis with the aforementioned estimate.

\subsubsection{Charged lepton channel at 250\,GeV}
\label{sec:results:ll250}
\noindent
The main SM background modes to the relevant signal process $e^+ e^- \to  Z h \to \ell^+ \ell^- b s$ arise from the processes  $e^+ e^- \to \ell^+ \ell^- \bar{b}b, \ell^+ \ell^- \bar{c}c$ and $\ell^+ \ell^- j j$ where the flavour of at least one final-state jet is mis-tagged and a $bj$ combination is identified. Two classes of processes contribute to this background, namely
\begin{description}[before={\renewcommand\makelabel[1]{\upshape\bfseries ##1}}]
	\item[Resonant Higgs processes] where the Higgs is produced in association with a leptonically decaying $Z$ boson, through the same topology as the signal.
	\item[Non-resonant processes] where the Higgs boson does not propagate in the Feynman diagrams, such as $ZZ$, $\gamma\gamma$ or $Z\gamma$ production.
\end{description}	
As mentioned before, resonant backgrounds are the most problematic ones since they exhibit the same kinematics as the signal process. Hence, they are only suppressed by flavour-tagging requirements.
Apart from the backgrounds listed above, charged-current processes like $e^+e^-\to\nu_\ell\ell sc$ could in principle also contribute, provided a second lepton that can arise for example from hadronic activity is detected.
However, defining suitable isolation criteria%
\footnote{We use a variant of the \texttt{DSiD} card that already implements electron, muon and photon isolation criteria. For details, see \cite{Potter2016b}.}
for the charged leptons and requiring them to have sufficiently large transverse momenta allows to reject this background very efficiently.

In order to discriminate the signal from the remaining backgrounds, we adopt the following set of cuts.
We ask for exactly one $b$ jet and one light-flavour jet $j$ which are supposed to both satisfy $p_T>\SI{10}{GeV}$ and $\absVal{\eta}<1.8$.
The $bj$ pair is required to have an invariant mass that fulfills $\invHiggs<\SI{30}{GeV}$.
We further ask for exactly one pair of same-flavour and opposite-sign isolated leptons (electrons or muons) with $p_T>\SI{10}{GeV}$ and $\absVal{\eta}<\num{2.44}$ each.
The lepton pair is additionally supposed to reconstruct the $Z$ boson mass within $\invZ<\SI{20}{GeV}$.
Finally, we ask the recoil mass, defined via $m^2_{\rm rec}:=s-2 \sqrt{s}E_{\ell \ell}+m_{\ell \ell}^2$, to be within a \SI{20}{GeV} mass window around the Higgs mass.

\begin{figure}[b]
	\centering
	\includegraphics[scale=0.83]{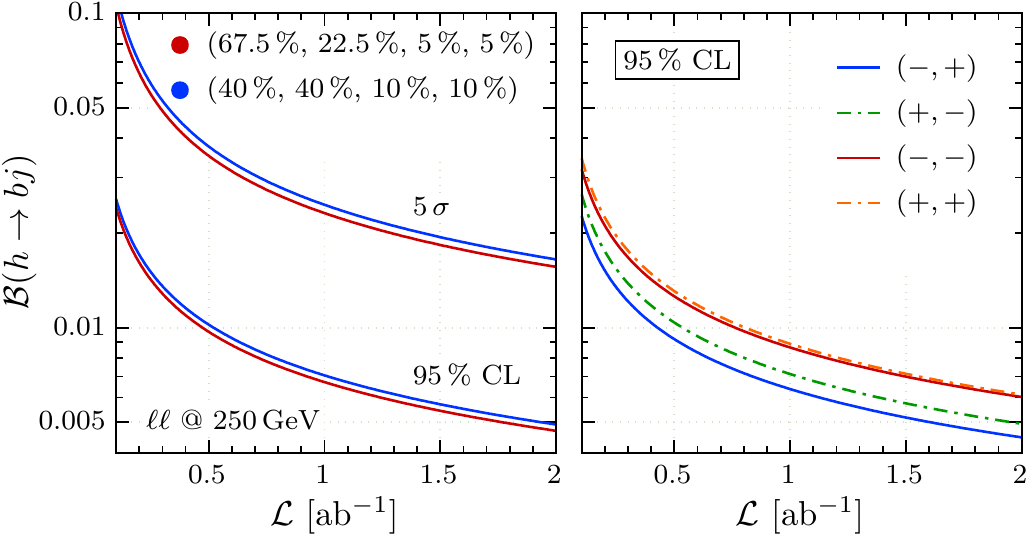}
	\caption{{\emph{Left}}: Expected \SI{95}{\%} \CL exclusion and $\SI{5}{\sigma}$ discovery reaches on \mbox{${\cal{B}}(h\to bj)$} in the charged lepton channel at $\sqrt{s}=\SI{250}{GeV}$ as a function of the total integrated luminosity for the two polarisation scenarios described in the text. {\emph{Right}}: Expected exclusion reaches for the four possible polarisation configurations individually.  Both plots assume $\epsilon_\text{syst}=\SI{1}{\%}$.}
	\label{fig:charg-lep-250}
\end{figure}

For these choices of selection cuts, the event yields for the charged lepton channel at $\sqrt{s}=\SI{250}{GeV}$ are reported in Tab.~\ref{tab:ll250:cutflow} assuming $\SI{2}{ab^{-1}}$ of accumulated data, $\br(h\to bs)\simeq \SI{0.73}{\%}$ for the signal and polarisation sharing of scenario 1.
The exclusion and discovery reach on $\br(h\to bj)$ as a function of the integrated luminosity with an assumed \SI{1}{\%} systematic background error are shown in Fig.~\ref{fig:charg-lep-250}.
The charged lepton channel will thus be able to exclude at \SI{95}{\%} \CL values of ${\cal{B}}(h\to b j)$ down to around $\SI{0.5}{\%}$, roughly one order of magnitude smaller than what can be obtained in the hadronic channel, with the ideal beam polarisation now being ${\cal{P_{-+}}}$.

\begin{table}[t]
	\centering
	\sisetup{round-mode=places,round-precision=0}
	\begin{tabular}{l | cccccc}
		\toprule		
	   & {Signal} & {$\ell\ell\bar{b}b$} & {$\ell\ell\bar{c}c$} & {$\ell\ell jj$} & {$\nu_\ell \ell s c$} & {$\nu_\ell \ell d u$} \\
		\colrule
Exp. & \num{225.435} & \num[round-precision=1]{1.241595e+05} & \num[round-precision=1]{1.396985e+05} & \num[round-precision=1]{4.0184e+05} & \num[round-precision=1]{5.23023e+06} & \num[round-precision=1]{5.233625e+06} \\
Jet tag. & \num{81.76875} & \num{4319.752} & \num{1903.621} & \num{2959.541} & \num{304.9528} & \num{151.6095} \\
$p_T^\ell$ & \num{79.197} & \num{3615.636} & \num{1551.644} & \num{2418.752} & \num{0} & \num{0} \\
$\invZ$ & \num{76.9885} & \num{2784.184} & \num{1051.278} & \num{1745.625} & \num{0} & \num{0} \\
$\recoilZ$ & \num{70.82825} & \num{1272.401} & \num{325.454} & \num{552.8228} & \num{0} & \num{0} \\
$\invHiggs$ & \num{50.55225} & \num{247.212} & \num{33.70475} & \num{66.0835} & \num{0} & \num{0} \\
		\botrule
	\end{tabular}
	\caption{
	Cutflow table for the signal and relevant backgrounds in the charged lepton channel at \SI{250}{GeV} with polarisation sharing scenario 1 and $\SI{2}{ab^{-1}}$ of integrated luminosity. For the signal we have assumed $\br(h\to bs)\simeq \SI[round-precision=2]{0.73}{\%}$.}
	\label{tab:ll250:cutflow}
\end{table}

\subsubsection{Neutrino channel at 250\,GeV and 500\,GeV}
\label{sec:results:vv}
\noindent
As outlined in Section \ref{sec:HiggsILC}, the neutrino channel with final state $e^+e^-\to \bar{\nu}\nu bs$ is interesting at both the \SI{250}{GeV} and the \SI{500}{GeV} runs.
The common search signature consists of exactly two jets plus missing transverse energy. Of the two jets, precisely one has to be $b$-tagged, while the other one must not receive any flavour tag.
Since we found that the relevant SM backgrounds for this signature as well as the most important cuts to isolate the signal are very similar for both energies, we discuss the two cases largely in parallel.

Four classes of SM processes contribute significantly to the background and are listed in the following.
\begin{description}[before={\renewcommand\makelabel[1]{\upshape\bfseries ##1}}]
	\item[Resonant Higgs processes] were already described in the beginning of Sec.~\ref{sec:results:leptonic} and make up the largest contribution because they are only suppressed by flavour-tagging requirements, but are otherwise irreducible.
	\item[Non-resonant processes] have the same final state as the resonant ones, but do not involve a Higgs decay. They may, for instance, originate from double-$Z$ production.
	\item[Quark pair production] ($e^+e^-\to\bar{b}b$, etc.) Uncertainties in the determination and/or reconstruction of the jets' four-momenta or other detector imperfections can lead to artificial missing energy. Although this background can be very efficiently reduced by a number of cuts (see below), it still contributes due to its large cross-section.
	\item[Charged-current processes] ($e^+e^-\to\nu_\ell\ell sc$ etc.)\\ arise \eg from double-$W$ production and contribute when the charged lepton is missed in the detector.
\end{description}

\begin{figure}[b]
    \centering
            \includegraphics[scale=0.83]{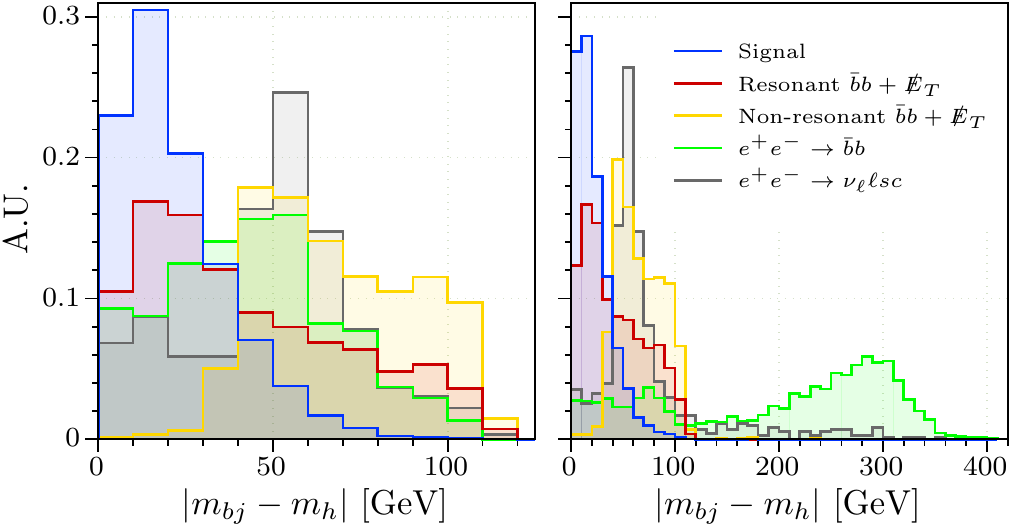}\hfill
    \caption{Normalised distributions of $\invHiggs$ after the 
   $\ETmiss$ cut for the neutrino channel at \SI{250}{GeV} (left) and \SI{500}{GeV} (right) with polarisation ${\cal{P}}_{-+}$ for both the signal and background processes.}
    \label{hadr:distr-neu}
\end{figure}

\begin{table*}[t]
	\centering
	\sisetup{round-mode=places,round-precision=0}
	\begin{tabular}{l|cccccc|cccccc}
		\toprule
		& \multicolumn{6}{|c}{$\sqrt{s}=\SI{250}{GeV}$} & \multicolumn{6}{|c}{$\sqrt{s}=\SI{500}{GeV}$} \\
		\colrule
		& {Signal} & {$(\bar{\nu}\nu)\bar{b}b$} & {$(\bar{\nu}\nu)\bar{c}c$} & {$(\bar{\nu}\nu)jj$} & {$\nu_\ell \ell s c$} & {$\nu_\ell \ell d u$} & {Signal} & {$(\bar{\nu}\nu)\bar{b}b$} & {$(\bar{\nu}\nu)\bar{c}c$} & {$(\bar{\nu}\nu)jj$} & {$\nu_\ell \ell s c$} & {$\nu_\ell \ell d u$} \\
		\colrule
		Exp. & \num{878.5} & \num[round-precision=1]{2.379695e+07} & \num[round-precision=1]{2.3969e+07} & \num[round-precision=1]{7.18989e+07} & \num[round-precision=1]{5.23023e+06} & \num[round-precision=1]{5.233625e+06} & \num{2490} & \num[round-precision=1]{7.50676e+06} & \num[round-precision=1]{8.1756e+06} & \num[round-precision=1]{2.29896e+07} & \num[round-precision=1]{3.59048e+06} & \num[round-precision=1]{3.60368e+06} \\
		Jet tag. & \num{434.551} & \num[round-precision=1]{1.020557e+06} & \num[round-precision=1]{5.153885e+05} & \num[round-precision=1]{8.62241e+05} & \num[round-precision=1]{2.368382e+05} & \num{66720.33} & \num{1146.983} & \num[round-precision=1]{1.623084e+05} & \num{98377.77} & \num[round-precision=1]{1.59813e+05} & \num[round-precision=1]{1.270692e+05} & \num{36040.1} \\
		No $\ell$ & \num{428.8582} & \num[round-precision=1]{9.689217e+05} & \num[round-precision=1]{5.111749e+05} & \num[round-precision=1]{8.507714e+05} & \num{7316.274} & \num{2429.642} & \num{1137.12} & \num[round-precision=1]{1.563268e+05} & \num{97544.72} & \num[round-precision=1]{1.572161e+05} & \num{4092.18} & \num{1539.002} \\
		$\ETmiss$ & \num{300.8683} & \num{21150.92} & \num{11827.73} & \num{13107.95} & \num{4399.931} & \num{1207.707}  & \num{1040.231} & \num{37246.96} & \num{20223.54} & \num{23285.07} & \num{3450.413} & \num{1216.682} \\
		$\invHiggs$ & \num{221.8598} & \num{4505.616} & \num{2419.87} & \num{2897.626} & \num{934.7593} & \num{402.1005} & \num{695.236} & \num{4617.385} & \num{567.729} & \num{812.703} & \num{287.738} & \num{116.371} \\
		$\DeltaPhi{b}{j}$ & \num{201.0008} & \num{1335.186} & \num{233.9653} & \num{201.8628} & \num{898.6252} & \num{375.3638} & \num{654.562} & \num{3945.501} & \num{299.127} & \num{522.079} & \num{270.103} & \num{107.382} \\
		\botrule
	\end{tabular}
	\caption{Cutflow table for the signal and backgrounds in the neutrino channel at \SI{250}{GeV} with $\SI{2}{ab^{-1}}$ and polarisation sharing of scenario 1 and \SI{500}{GeV} with $\SI{4}{ab^{-1}}$ and polarisation sharing of scenario 2. For the signal we have assumed $\br(h\to bs)\simeq \SI[round-precision=2]{0.73}{\%}$.}
	\label{tab:vv250:cutflow}
\end{table*}

\begin{figure*}[t]
	\centering
	\subfloat[][]{\includegraphics[scale=0.83]{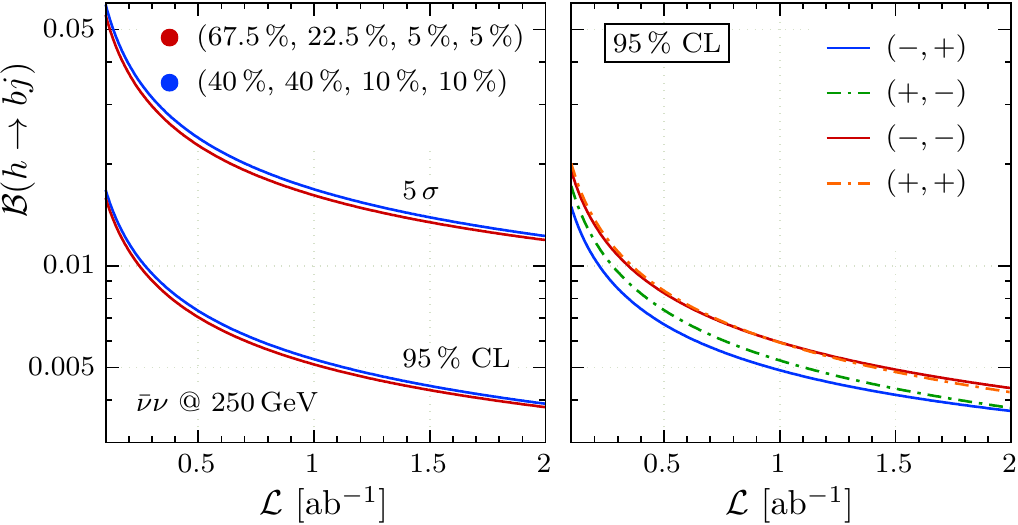}}
	\hfill
	\subfloat[][]{\includegraphics[scale=0.83]{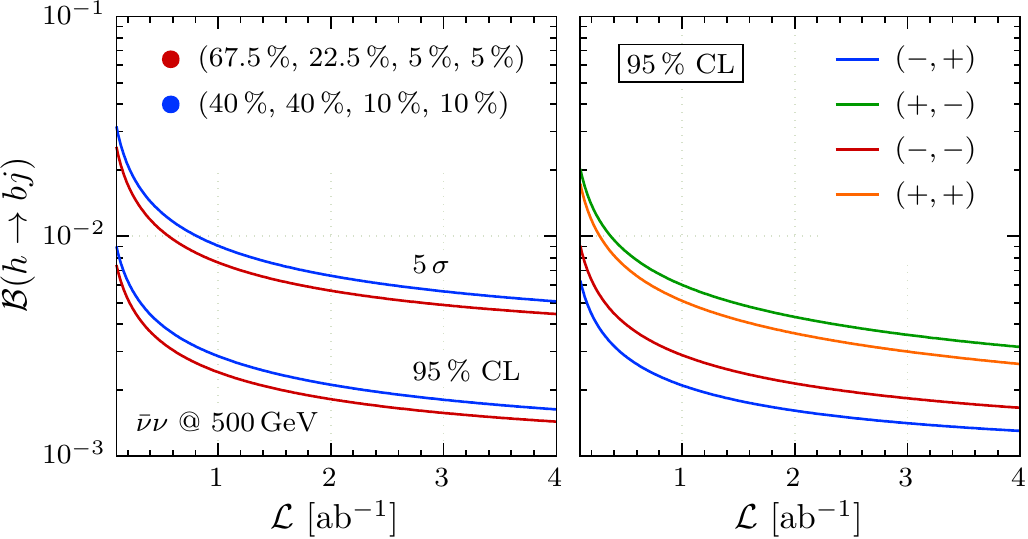}}
	\caption{{\emph{(a) Left}}: Expected \SI{95}{\%} \CL exclusion and $\SI{5}{\sigma}$ discovery reaches on \mbox{${\cal{B}}(h\to bj)$} in the neutrino channel at $\sqrt{s}=\SI{250}{GeV}$ as a function of the total integrated luminosity for the two polarisation scenarios described in the text. {\emph{(a) Right}}: Expected exclusion reaches for the four possible polarisation configurations individually.  Both plots assume $\epsilon_\text{syst}=\SI{1}{\%}$.
	{\emph{(b)}}: Same as \textit{(a)} but for $\sqrt{s}=\SI{500}{GeV}$.}
	\label{fig:neu-250}
\end{figure*}

All but the resonant processes do not involve Higgs decays, such that they can be efficiently rejected by the two-jet invariant mass cut, $\invHiggs$, see Fig.~\ref{hadr:distr-neu}.
Furthermore, requiring a significant amount of missing transverse energy helps to reduce the background from events with less than two neutrinos.
We select signal events requiring exactly one $b$-tag jet and one light-flavour jet with $p_T>\SI{10}{GeV}$ and $\absVal{\eta}<1.8$. We then veto on the presence of reconstructed isolated charged leptons. Further we require $\invHiggs<\SI{30}{GeV}$ and  $\ETmiss>\SI{35}{GeV}$ for $\sqrt{s}=\SI{250}{GeV}$ and $\invHiggs<\SI{25}{GeV}$ and  $\ETmiss>\SI{25}{GeV}$ for $\sqrt{s}=\SI{500}{GeV}$. We finally require that the azimuthal distance between the two jets satisfies $\DeltaPhi{b}{j}<3.0$, which helps to reduce the background contribution from quark-pair final states for which the jets are mainly back-to-back. Note that the $\ETmiss$ cut slightly differ for the two considered centre-of-mass energies. This is primarily caused by the differences in available energy as well as by the different kinematics of the two main production channels.

For these choices of selection cuts the event yields for the signal, with $\br(h\to bs)\simeq \SI{0.73}{\%}$, and the SM backgrounds are reported in Tab.~\ref{tab:vv250:cutflow}.
For the $\SI{250}{GeV}$ run, polarisation sharing scenario 1 and $\SI{2}{ab^{-1}}$ of accumulated data were assumed, whereas we adopted scenario 2 and an integrated luminosity of $\SI{4}{ab^{-1}}$ for $\sqrt{s}=\SI{500}{GeV}$.
The corresponding exclusion and discovery reach on ${\cal{B}}(h\to bj)$ as a function of the integrated luminosity for a \SI{1}{\%} systematic background error are displayed in Fig.~\ref{fig:neu-250}.
The results show that the neutrino channel at $\sqrt{s}=\SI{250}{GeV}$ is able to set a limit on \mbox{$\br(h\to bj)$} slightly better than that obtained from the charged lepton channel at the same centre-of-mass energy.
We finally see, that the best sensitivity of all search modes under consideration is offered by the neutrino channel at the \SI{500}{GeV} run. To be precise, a \SI{95}{\%} \CL upper limit on \mbox{${\cal B}(h\to bj)$} of approximately $\SI{0.2}{\%}$ can be set at the end of the ILC operations.
The optimal beam polarisation for the neutrino channels is found to be ${\cal{P_{-+}}}$.

Let us finally compare the obtained result of $\br_{95\%} \approx \num{2e-3}$ to the estimate in Eq.~\eqref{eq:bkgdEstimate:br95}.
In a benchmark scenario with \mbox{$\br_0=\SI{0.73}{\%}$} and at a full \SI{500}{GeV} run with only ${\cal{P_{-+}}}$ beams, we expect a total of \num{5256} signal events to be produced.
Factoring in the assumed tagging efficiency from Table \ref{tab:bkgdEstimate}, this number reduces to $S_0=\num{3742}$, giving a maximally achievable sensitivity of $\br_{95\%} \gtrsim \num{5e-4}$.
Additionally taking into account particle reconstruction efficiencies, detector acceptance and a more realistic polarization sharing (but no further kinematical cuts) one ends up with $S_0=\num{1137}$ and $\br_{95\%} \gtrsim \num{9e-4}$.
Note that in a realistic analysis, additional cuts would have to be applied in order to reduce non-resonant background contributions.
The above estimate is already quite close to our full result thus showing that the performance of the neutrino channel is indeed limited by the resonant background modes and flavour tagging capabilities.


\section{Summary and conclusions}
\label{sec:conc}
\noindent
In this paper we have discussed the potential of a future $e^+e^-$ collider in directly searching for quark flavour-violating Higgs decays involving a bottom quark and a light-flavour quark.
As a benchmark machine, we chose the planned International Linear Collider (ILC).
Focusing on the two main Higgs production modes at the ILC \textendash\ the Higgs-strahlung and the $W$-fusion processes \textendash\ we identified the most promising analysis channels for the $\SI{250}{GeV}$ run to be the ones involving a pair of neutrinos or charged leptons in the final state.
Adopting standard ILC beam polarisation scenarios, we found that a dedicated analysis is expected to exclude at \SI{95}{\%} \CL exotic Higgs branching ratios of the order of $\SI{0.5}{\%}$ with $\SI{2000}{fb^{-1}}$ of integrated luminosity, while a 5$\sigma$ significance can be attained for order $\SI{1}{\%}$ decay rates.
In contrast, the hadronic final state will have a sensitivity reduced by a factor of ten due to the large combinatorial four-jet background. 
\begin{figure}[t]
	\centering
	\includegraphics[scale=0.8]{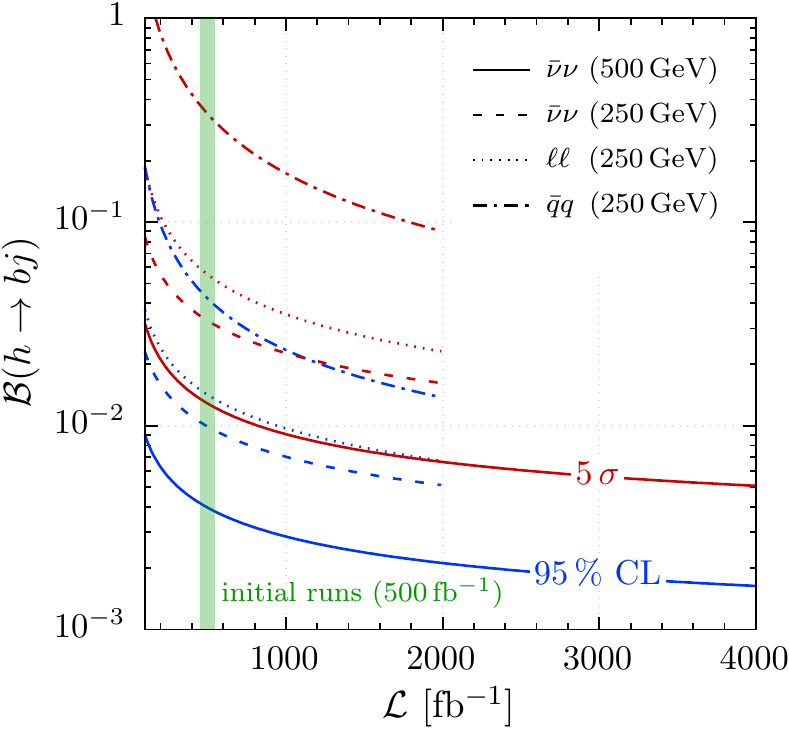}
	\caption{
Comparison between the expected \SI{95}{\%} \CL exclusion (blue) and $\SI{5}{\sigma}$ discovery (red) reaches on ${\cal{B}}(h\to bj)$ for the various channels as a function of total integrated luminosity specific for the given centre-of-mass energy.
The polarisation sharing is scenario 1 and 2 for the channels at $\sqrt{s}=\SI{250}{GeV}$ and $\SI{500}{GeV}$, respectively. The plot assumes $\epsilon_\text{syst}=\SI{1}{\%}$.
}
	\label{fig:final}
\end{figure}

At a centre-of-mass energy of $\SI{500}{GeV}$ with $\SI{4000}{fb^{-1}}$ of accumulated data and by exploiting the final state with a pair of neutrinos, an upper limit on \mbox{$\mathcal{B}(h\to bj)$} of approximately $\SI{0.2}{\%}$ is expected, while a discovery will be possible if $\mathcal{B}(h\to bj)$ is greater than $\SI{0.5}{\%}$.
Even the initial run with only $\SI{500}{fb^{-1}}$ of integrated luminosity will be able to test, at \SI{95}{\%} CL, branching ratios as small as \SI{0.4}{\%}.
In particular, we demonstrated that our results are close to the fundamental bound on the sensitivity, which is determined by resonant background rates and flavour tagging capabilities only.

The outcome of our study is then summarised in Fig.~\ref{fig:final} where both the $\SI{95}{\%}$ CL exclusion (blue) and the $\SI{5}{\sigma}$ discovery (red) reaches on $\mathcal{B}(h\to bj)$ are illustrated for the four channels analysed in the paper showing only the polarisation scenario and integrated luminosity relevant for a given centre-of-mass energy.

In conclusion, our analysis has shown that the prospects for directly testing exotic Higgs decays at the ILC are indeed promising at both of its planned centre-of-mass energies, and that the projected limits from the full \SI{500}{GeV} run are in the same ballpark as the ones that can be obtained from low-energy flavour measurements of the Wilson coefficient of the operators mediating flavour-changing neutral currents in the Higgs sector. These experiments provide, however, indirect probes of quark flavour-violating couplings to which direct tests at colliders are complementary.



\section*{acknowledgements}
\noindent
The authors would like to thank Wolfgang Kilian and Thorsten Ohl for help regarding the use of {\tt Whizard} and Nishita Desai for clarifications on {\tt PYTHIA8}.\\
A.H.~acknowledges support by the IMPRS-PTFS.

\bibliographystyle{references/bibstyle}
\bibliography{references/References-MPIK}

\end{document}